\documentclass[aps,epsfig,showpacs]{revtex4}
\usepackage{amsmath}
\usepackage{amssymb}
\usepackage{psfrag}
\usepackage[dvips]{epsfig}
\usepackage{latexsym}
\usepackage[dvips,figuresright]{rotating}

\begin{document}
\title{Exotic trees}

\author{
Z. Burda$^{1,2}$, J. Erdmann$^1$, 
B. Petersson$^1$ and M. Wattenberg$^1$}

\affiliation{
\vspace{0.4cm}
$^1$ Fakult\"at f\"ur Physik, Universit\"at Bielefeld, P.O.Box 100131, 
D-33501 Bielefeld, Germany. \\
$^2$ Institute of Physics, Jagellonian University, 
ul. Reymonta 4, 30-059~Krak\'ow, Poland. \\
}

\begin{abstract} \noindent
We discuss the scaling properties of free  branched polymers.
The scaling behaviour of the model is classified by the Hausdorff 
dimensions for the internal geometry~: $d_L$ and $d_H$, 
and for the external one~: $D_L$ and $D_H$. 
The dimensions $d_H$ and $D_H$ characterize
the behaviour for long distances 
while $d_L$ and $D_L$ for short distances.
We show that the internal Hausdorff dimension is 
$d_L=2$ for generic and scale-free trees,
contrary to $d_H$ which is known be equal two for generic
trees and to vary between two and infinity for scale-free
trees. We show that the external Hausdorff dimension
$D_H$ is directly related to the internal one as 
$D_H = \alpha d_H$, where $\alpha$ is the stability index 
of the embedding weights for the nearest-vertex 
interactions. The index is $\alpha=2$ 
for weights from the gaussian domain of attraction 
and $0<\alpha <2$ for those from the L\'evy 
domain of attraction. 
If the dimension $D$ of the target space is larger 
than $D_H$ one finds $D_L=D_H$, or otherwise $D_L=D$.
The latter result means that the fractal structure 
cannot develop in a target space which has too low dimension.
\end{abstract}

\maketitle

\section*{Introduction}
In recent years the theory of random geometry \cite{quageo} has
become a powerful method of investigating problems
in many areas of research ranging from the 
statistical theory of membranes \cite{memb1,memb2}, 
branched polymers \cite{poly1,poly2}
and complex networks \cite{net1,net2} 
to fundamental questions in string theory \cite{str1,str2,str3} 
and quantum gravity \cite{grav1,grav2,grav3}.

Those problems have in common
that they can be described by a
dynamically alternating geometry which 
undergoes fluctuations of a statistical or quantum 
nature. The dynamics of such fluctuations can
be modelled using the concepts of the statistical 
ensemble and the partition function in a similar way 
as it is done in particle physics by the methods of
lattice field theory. 

Contrary to lattice field theory where
the partition functions run over field 
configurations on a rigid geometry,
the geometry itself is variable here.
Since the geometry is dynamical 
many new features
occur like for instance geometrical 
correlations or the influence of the random
geometry on the fields living on it. 

Similarly as in field theory where the concepts of universality, 
critical exponents, correlations, {\em etc} are 
independent of whether one discusses a field theoretical 
model of magnetism or a quantum theoretical model of particles, 
also in the theory of random geometry many questions
are independent on details and may be addressed using
general methods. General concepts can be best developed 
on an analytically treatable model. 
In field theory the role of a 
test bed is played by the Ising model while
in random geometry by the branched polymer model 
\cite{bp1,bp2,bp3,bp4}.

Despite its simplicity 
the branched-polymer model has
a rich phase structure exhibiting different
scaling properties of the fractal geometry 
and the correlation functions. 

The model has internal and external geometry 
sectors similarly as the Polyakov string \cite{str1}. 
As the Polyakov string, 
it can also be interpreted as a model 
for quantum objects 
embedded into a $D$-dimensional 
target space or a model of quantum gravity interacting
with $D$-scalar fields. 
The term quantum gravity refers to 
an Euclidean Feynman integral expressed
by a sum over diagrams representing 
the nearest-neighbour 
relations between points of a discrete set. 
In more realistic models the sum runs
over higher dimensional simplicial manifolds
and can be interpreted as a regularized Feynman
integral over Riemannian structures 
on the manifold \cite{str2,str3,grav1}.
The intuition which one can gain from the 
analytic solution of the branched-polymer model 
is very helpful for considerations of more
complicated models. In fact, the model
has proven already many times to be extremly useful
to test and develop various ideas concerning
random geometry \cite{bp1}-\cite{ft1}.

In addition to this general interest in this
model there is a specific motivation which
is related to the reduced super-symmetric Yang-Mills 
matrix model \cite{sym}. This model was introduced 
as a non-perturbative definition for superstrings \cite{ikkt1}
and referred to afterwards as the IKKT model.
The one-loop approximation 
of this model leads to a model of graphs which 
have as a back-bone a branched-polymer 
with power-law weights for the link lengths .
The IKKT model is believed to provide a dynamical
mechanism for the spontaneous breaking of the Lorentz
symmetry from ten to four dimensions \cite{ikkt2}. 
If one tries to understand the breaking in terms of the one-loop
level approximation one finds it to be 
related to the fractal properties of 
the branched-polymers, which have the Hausdorff 
dimension equal to four \cite{ikkt2,bpj}. 
The question of the spontanous symmetry breaking was 
investigated also by many other methods with help of
which one was able to gain an insight into underlying
physical mechanisms \cite{ssb1}-\cite{ssb8}. 

In the simplest case one considers generic trees
with the nearest-vertex interactions given by
gaussian weights or weights which belong to the gaussian
universality class \cite{poly1,poly2}. In this paper we also discuss
the nearest-vertex interactions, but we 
extend the discussion to non-gaussian weights
in particular to power-law weights \cite{levy1}.
The energetical costs to generate long 
links on the polymer is then much smaller than for 
gaussian ones. In the extremal situation, 
when the power-law exponent $\alpha$ of the
link length distribution $\sim x^{-1-\alpha}$
lies in the interval $\alpha \in (0,2)$,
very long links are spontanously generated
on the tree and their presence shifts the  
model to a new universality class which 
can be called the class of L\'{e}vy branched polymers 
that similarly as L\'{e}vy paths exhibit a new 
scaling behaviour. 

The scaling properties and the universality class of the
model depends also on the internal 
branching weights of the trees \cite{ft1}. Under a
change of the weights the 
model may undergo a transition
from the phase of elongated trees with the
internal Hausdorff dimension $d_H=2$, known
as generic trees, 
to the phase of collapsed trees with $d_H=\infty$
which are localized around a singular vertex 
of high-connectivity \cite{ft2,ft3}. In between there is a phase
of scale-free trees which may have any Hausdorff dimension
between $d_H=2$ and $d_H=\infty$ \cite{net2,ft1}. We show that
the internal properties decouple from the embedding in the
target space but on the other hand that they strongly
affect the embedding~: the external Hausdorff dimension
is proportional to the internal one $D_H = \alpha d_H$.

In this paper we cover the entire classification of the
universality classes of branched polymers with the 
nearest-vertex interactions. We hope it can be treated 
as a starting point in the discussion of models with 
more complicated interactions like for example those 
with correlations between neighbouring links,
or self-avoidance in the embedding space.

Many pieces of this classification have been 
discussed for the gaussian branched-\-polymers already
\cite{poly1,poly2,net2,bp2,ft1,bpj}. Several well-known 
results have been summarized within the appendix 
in which we present a systematical treatment of 
the internal geometry in terms of generating functions. 

The extension of this classification 
to weights with power-law tails and to the case 
when the internal geometry is non-generic, 
is presented in the main-stream of the text.
Emphasis is put on calcuations of the two-point
functions. Throughout the paper we also
stress the factorization property of the
internal and external geometry, which
allows us to clearly separate the discussion
of the internal geometry before considering 
the entire model. It also permits us to reveal
many interesting relations between the correlation
functions of the external space to those for
the internal geometry. 

\section*{The model}
We consider a canonical ensemble of 
trees embedded in a $D$-dimensional target space. 
The partition function of the ensemble is defined as
a weighted sum over all labeled trees with $N$ vertices.
The set of such labeled trees, containing $N^{N-2}$ elements, 
will be denoted by ${\cal T}_{N}$. 
The statistical weight of a tree is given by
a product of an internal weight $W_T$ which depends only 
on the internal geometry of the tree, and an external one which 
depends on the positions of the (tree) vertices 
in the target space. We shall consider trees
with nearest-neighbour interactions 
for which the partition function reads~:
\begin{equation}
Z_N = \frac{1}{N!} \sum_{T\in {\cal T}_N} W_T 
\int \prod_{i=1}^{N} {\rm
d}^D x_i \ \prod_{\langle
jk \rangle} 
f\left(\vec{r}_{jk} \right)
\: .
\label{zn}
\end{equation}
The external weight of a tree is a product of
link weights $f(\vec{r}_{jk})$ 
which depend exclusively on the link
vector $\vec{r}_{jk} \! = \! \vec{x}_{j} - \vec{x}_{k}$. 
Saying alternatively,
the energy cost of the embedding 
of the tree in the target space is
a sum of energy costs 
of the independent embedding of links. The
second product in (\ref{zn})
runs over the set of (unoriented) linked
vertex pairs denoted 
by $\langle jk \rangle$.

The most natural choice of the internal weights
is $W_T=1$. We could entirely stick to this choice of weights,
but since we want to discuss the problem of universality 
we also want to check whether a modification of the
weights will change the scaling properties and hence
the universality \cite{ft1}.

Here we will restrict our considerations to internal weights 
which can be written as a product of weights $w_{q}$
for the individual vertices~:
\begin{equation}
W_{T} = w_{q_{1}} w_{q_{2}} \cdots w_{q_{N}} \: .
\label{weightdef}
\end{equation}
Each vertex weight only depends on the
degree of the vertex that means the number of links emerging
from it. The internal properties of the 
model are determined when the whole set of
branching weights $\{ w_q \}$ for $w=1,2,\dots,\infty$
is specified. We demand that~:
\begin{equation}
w_1 >0 \ , \quad w_q\ge 0 \ , \quad w_s > 0 
\label{wcond}
\end{equation}
for all $q=2,\dots,\infty$ and at least one
$s>2$. If $w_1$ were zero $W_T$ would vanish for
all tree graphs while if all $w_q$ for $q>2$ 
were zero then the weights $W_T$ would 
vanish for all trees except chain-structures.

Note that the model is invariant with respect
to translations in the external space~: 
$\vec{x}_i \rightarrow
\vec{x}_i + \vec{\delta}$. 
Because of the
translational zero-mode the
partition function (\ref{zn}) is infinite.
One can make it finite by dividing
out the volume
$V = \int {\rm d}^D x$ of the translational zero-mode~:
\begin{equation}
z_N = \frac{Z_N}{V} \: .
\label{zeromode}
\end{equation}
This can for example be realized by fixing the position
of the center of mass of the trees.

Trees which can be obtained from 
each other by a permutation of the vertex labels
contribute with 
the same statistical weight. For a tree with
$N$ vertices
there are $N!$ such vertex permutations. 
In order to avoid overcountings
one introduces the standard factor $1/{N!}$ to 
the definition of the partition function (\ref{zn}). 
This factor divides out the volume of the
permutation group of the vertex labels.
The number of all labeled trees
counted with this factor
$N^{N-2}/N! \sim N^{-5/2} e^{N}$ is 
exponentially bounded 
in the $N \rightarrow \infty$ limit.
If one defines the grand-canonical partition function 
\begin{equation}
{\cal Z} = \sum_{N=2}^{\infty} z_{N} g^{N} = \sum_{N=2}^{\infty} z_{N} e^{- \mu N}\: , 
\label{grandz}
\end{equation}
one can see that it is well defined as long as 
the fugacity $g$ is smaller than the radius of 
convergence of the series, which in the particular
case $W_{T}=1$ is 
equal $g_0=e^{-\mu_0} = e^{-1}$. More generally,
as long as $z_N$ grows only exponentially 
for large $N$ the grand-canonical partition 
function has a non-vanishing 
radius of convergence and hence  
one can savely define ${\cal Z}$. 

The statistical average of a quantity $Q$ 
defined on the ensemble (\ref{zn}) is given by~:
\begin{equation}
\langle Q \rangle_N \equiv \frac{1}{z_N} 
\frac{1}{N!} \sum_{T\in {\cal T}_N} W_T 
\int \prod_{i=1}^{N} {\rm d}^D x_i \ \prod_{\langle jk \rangle}
f\left(\vec{r}_{jk} \right) \: Q \: .
\label{an}
\end{equation}
For translationally invariant quantities the averages
are proportional
to the volume $V$ of the translational zero-mode. 
For such quantities
one should rather speak of an average 
density per volume element of
the target space~:
$\langle Q \rangle_N/V$ which is a finite number.
In
particular $\langle 1 \rangle_N/V = 1$.

We will frequently distinguish
between 
the internal and external geometry of the trees. 
By the former
we mean the connectivity of the corresponding
tree graph, by the latter
its embedding in the external space. 
For example, the internal (geodesic)
distance
between two vertices $i$ and $j$ of a graph is defined as
the
number of links of the shortest path connecting them, while the
external
distance is given by the length of the vector 
$\vec{x}_{i}-\vec{x}_{j}$.
Note that the path between 
$i$ and $j$ is unique for tree graphs. Thus
the length of this 
path, i.e. the number of its links, determines
the internal geodesic distance. 

The properties of the embedding in the
external space depend on the link weight 
function $f(\vec{r})$ (see (\ref{zn})).  We
consider isotropic weights depending only 
on the link length.  That
means $f(\vec{r}) = f(r)$, where $r=|\vec{r}|$. 
We further assume that $f(\vec{r})$ is a positive 
integrable function. Without loss of generality,
we can then choose the normalization to be~: 
$\int {\rm d}^D r \ f(\vec{r}) = 1$. This allows us 
to interprete $f(\vec{r})$ as a probability density.

\section*{Correlation functions}

The fundamental quantities which encode the information
about the statistical properties of the system
are the correlation functions. 
For the canonical ensemble
with the partition function (\ref{zn}), the
$m$-point correlation functions are defined as~:
\begin{equation}
G^{(m)}_N(\vec{X}_{A_1},\dots
,\vec{X}_{A_m}) \equiv
\left\langle {\prod_{k=1}^{m}} {}^{{}^\prime} 
\frac{1}{N} \sum_{a_k=1}^N
\delta (\vec{x}_{a_k} - \vec{X}_{A_k})
\right\rangle_N \: ,
\label{gNn}
\end{equation}
where the brackets $\left\langle \: . \: \right\rangle_N$
on the right hand side denote the average
over the ensemble (\ref{an}).
If one multiplies all sums in the product in
(\ref{gNn}) one obtains a sum of terms being
products of delta functions. The prime in 
equation (\ref{gNn}) means that all terms which
contain two or more identical delta functions are
skipped from this sum. This exclusion principle
applies only to the situation when any two
arguments of $G^{(m)}_{N}(\vec{X}_{A_1},\dots,\vec{X}_{A_m})$
are identical.

What we will show now is that the problem of
determining the $m$-point correlation function 
can be devided into two sub-problems. The first step
is to determine the internal two-point function.
This can in general be done independently of
a particular choice of the embedding weights 
$f(\vec{r})$. The second is to use the information
encoded in the internal two-point function to
determine the external properties of the trees.

In order to see that the internal geometry
of the trees does not depend on the choice 
of the link weight function,
consider a tree
graph and calculate the following two integrals for this tree~:
\begin{eqnarray}
\label{t1}
\int \prod_{i=1}^{N} {\rm d}^D x_i \ \prod_{\langle jk \rangle}
f\left(\vec{r}_{jk} \right) \cdot \delta(\vec{x}_a - \vec{X}_A) &=& 1 \:, \\
\label{t2} \int \prod_{i=1}^{N} {\rm
d}^D x_i \ \prod_{\langle jk \rangle} f\left(\vec{r}_{jk} \right) \
\cdot \ \delta(\vec{x}_a - \vec{X}_A) \
\delta(\vec{x}_b - \vec{X}_B) &=& f_n ( \vec{X}_B-\vec{X}_A ) .
\end{eqnarray}
The first integral corresponds to the embedding weight factor
for a tree whose $a$-th vertex is fixed at the position
$\vec{X}_A$ in the target space. Since the model is translationally
invariant the result of the integration does not depend on the position.
This result can be obtained by changing the integration variables
from position vectors of all vertices $i\ne a$ to
link vectors $\vec{r}_{jk} = \vec{x}_j - \vec{x}_k$ 
for which the integration completely factorizes.
The Jacobian for such a change of the integration variables is equal
one.

The second integral (\ref{t2}) gives the weight factor for a tree whose
vertices $a$ and $b$ are fixed at the positions $\vec{X}_A$
and $\vec{X}_B$ in the external space. The result 
depends only on the difference $\vec{X} \equiv \vec{X}_B-\vec{X}_A$
 and the number of links, $n$, of 
the path connecting the vertices $a$ and $b$. If one now
changes the integration variables from vertex positions 
to link vectors, as before, one can see 
that all integrations, except those for links on the path
between $a$ and $b$, factorize. The sum of the link vectors
on the path is restricted to $\vec{X} = \vec{X}_B-\vec{X}_A$.
If we label the links of the path by consecutive numbers
from $1$ to $n$, we can write~:
\begin{eqnarray}
f_n(\vec{X}) &=& \int \prod_{i=1}^n
\left( {\rm d}^D r_i f(\vec{r}_i) \right) \, 
\delta\left(\sum_{a=1}^n \vec{r}_a - \vec{X}\right)
\nonumber \\ \label{convo} &=& \int \frac{{\rm
d}^D p}{(2\pi)^D} \left[\widehat{f}(\vec{p})\right]^n 
\ e^{- i \vec{p} \vec{X}} \: ,
\end{eqnarray}
where $\widehat{f}(\vec{p})$ is the characteristic function
of the probability distribution $f(\vec{r})$~:
\begin{equation}
\widehat{f}(\vec{p}) \equiv \int {\rm d}^D r \ f(\vec{r}) \
e^{i\vec{p}\vec{r}} = \langle e^{i\vec{p}\vec{r}} \rangle_f \: .
\label{charact}
\end{equation}
It is important that the results of the integrations 
(\ref{t1}) and (\ref{t2}) do not depend on 
the internal geometry of the underlying tree graph.
In particular, using (\ref{t1}) and (\ref{t2}) we find
the partition function (\ref{zeromode}), the
one-point and two-point correlation functions to reduce
to the following form~:
\begin{eqnarray}
\label{izn} z_N &=& \frac{1}{N!} \sum_{T\in {\cal T}_N} W_T \: , \\
\label{G1} G^{(1)}_N(\vec{X}_A) &=&1 \: ,\\
\label{G2} G^{(2)}_N(\vec{X}_A,\vec{X}_B) &=& 
G^{(2)}_N(\vec{X}) =
\sum_{n=1}^{N-1} f_n(\vec{X}) \ g^{(2)}_N(n) \: .
\label{Gfg}
\end{eqnarray}
We have denoted the (canonical) two-point function of the internal 
geometry by $g^{(2)}_N(n)$~:
\begin{equation}
\label{g2}
g^{(2)}_N(n) = \frac{1}{z_N} \frac{1}{N!} \sum_{T \in {\cal T}_N} W_T
\left(\frac{1}{N^2} \sum_{a,b\in T} \delta_{|a-b|,n} \right) \: .
\end{equation}
It is normalized to unity~: $\sum_{n=0}^{N-1} g^{(2)}_{N}(n) = 1$.
The normalized internal two-point function 
gives us the probability that two 
randomly chosen vertices on a random tree of size $N$ are 
separated by $n$ links.
In the last formula the geodesic internal distance between 
the vertices $a$ and $b$ is denoted by $|a-b|$. 

The expressions (\ref{izn}) for $z_N$, (\ref{G1}) 
for $G^{(1)}_N(\vec{X})$ and (\ref{g2}) 
for $g^{(2)}_{N}(n) \:$ are independent of the link weight
factor $f(\vec{r})$. Thus 
the properties of the internal geometry, as mentioned, 
can be considered independently of and prior to the embedding.

The external two-point function 
$G^{(2)}_N(\vec{X})$ can be interpreted
as the probability density for two random vertices on a tree of size
$N$ to be embedded in the external space with the relative position
$\vec{X} = \vec{X}_B - \vec{X}_A$. The probability normalization condition
reads~:
\begin{equation}
\int {\rm d}^D X_B \: G^{(2)}_{N}(\vec{X}_A,\vec{X}_B)
= G^{(1)}_{N}(\vec{X}_A) = 1 \, .
\end{equation}
The right hand side of (\ref{G2}) can be understood as
a conditional probability. First, we choose two random
vertices on a tree of size $N$ and calculate
the probability $g^{(2)}_N(n)$ that there 
are $n$ links on the path connecting them. For
this path, which is a random path in the embedding
space consisting of $n$ links, we can calculate 
the probability (density) $f_n(\vec{X})$ 
that its endpoints are located 
with the relative position $\vec{X} = \vec{X}_B-\vec{X}_A$. 
Since the internal geometry decouples from
the external one, the probability densities $f_n(\vec{X})$ 
and $g^{(2)}_N(n)$ are independent of each 
other and can hence be calculated separately.

Similarly, higher correlation functions can be obtained
from  the corresponding internal correlation functions. 
For example, the three-point correlation function is~:
\begin{eqnarray}
G^{(3)}_N(\vec{X}_A,\vec{X}_B,\vec{X}_C) &=& \sum_{n_a,n_b,n_c} 
g^{(3)}_N(n_a,n_b,n_c) \times \label{G3} \\ 
&\times& \int {\rm d}^D X \,
f_{n_a}(\vec{X}_A-\vec{X}) 
f_{n_b}(\vec{X}_B-\vec{X}) 
f_{n_c}(\vec{X}_C-\vec{X}) \nonumber
\: .
\end{eqnarray}
The three paths $ab$, $bc$ and $ac$ between the  
vertices $a$, $b$ and $c$ of the tree 
can be decomposed into three
pieces, namely $am$, $bm$, $cm$ between them and the common middle
point $m$. The summation indices 
$n_a$, $n_b$, $n_c$ denote the internal lengths of these pieces,
and $\vec{X}$ denotes the position of the common vertex 
in the external space. 
The internal three-point function then reads~:
\begin{eqnarray}
\label{g3}
g^{(3)}_N(n_a,n_b,n_c) &=& \frac{1}{z_N} \frac{1}{N!} 
\sum_{T \in {\cal T}_N} W_T \times \\
&\times& \left(\frac{1}{N^3} \sum_{a,b,c\in T} 
\delta_{|a-b|,n_a+n_b} \delta_{|b-c|,n_b+n_c} \delta_{|a-c|,n_a+n_c} \right) \nonumber \: .
\end{eqnarray}
One could extend this construction further.

Note that the most important piece of information
is already encoded in the two-point function
and is inherited by the higher correlation
functions.  In fact, one can directly derive the 
higher correlation functions from the two-point 
function, using a simple composition rule for the
tree graphs which enormously simplifies in the grand canonical ensemble.
In the next section we shall thus concentrate on the two-point function.

\section*{Fractal geometry}

The canonical two-point correlation functions $g^{(2)}_N(n)$ and
$G^{(2)}_N(\vec{X})$ contain the information about
the fractal structure of the internal and external geometry,
respectively. The average distance for the internal
geometry, given by the average number of links
between two vertices on the tree, is the first moment 
of the probability distribution (\ref{g2})~:
\begin{equation}
\label{naverage}
\langle n \rangle_N = \sum_n n \: g^{(2)}_N(n)
\, .
\end{equation}
One expects the following scaling behaviour for large $N$~:
\begin{equation}
\langle n \rangle_N \sim N^{1/d_H} \, .
\label{ddH}
\end{equation}
The exponent $d_H$ relates the systems average (internal) 
extent $\langle n \rangle_{N}$ to its size $N$ and is thus
called the internal Hausdorff dimension. This exponent controls the
behaviour for large distances growing with the system
size $N$.  One can also introduce a local definition of the fractal
dimension for distances in the scaling window 
$1\ll n \ll N^{1/d_H}$. The scaling window contains distances
between the scale of the ultraviolet cut-off and below
the infrared scale set by the system size. 
This is a sort of thermodynamic definition which becomes valid 
locally for sufficently large $N$.  
In a large system one can be 
interested in how the volume of a local
ball (or sphere) depends on its radius $n$. The volume of the
sphere can be calculated as the number of vertices lying
$n$ links apart from a given vertex~:
\begin{equation}
g^{(2)}_{N \rightarrow \infty}(n) \sim n^{d_L-1} \quad {\rm for} \quad
1 \ll n\ll N^{1/d_H} \: .
\end{equation}
The definition $d_L$ is more practical for a local observer, for 
example, someone who lives in a fractal geometry and 
wants to determine its fractal dimension.
The global definition $d_H$ is accessible only for
an observer who can survey the whole system from 
outside \cite{foot1}.

In a similar way we can define the external Hausdorff dimension.
In order to do this we first have to introduce a measure of 
the systems extent in the external space. Such a measure is 
provided by the gyration radius~:
\begin{equation}
R^2 = \frac{1}{ N^2} \sum_{i,j} (\vec{X}_i
- \vec{X}_j)^2 = \frac{2}{N} \sum_{i} (\vec{X}_i - \vec{X}_{CM})^2 \, ,
\label{gyrr}
\end{equation}
where $\vec{X}_{CM} = \sum_i \vec{X}_i/N$ is the target space 
position of the systems center of mass.
The statistical average of the gyration radius is
directly related to the two-point function, namely~:
\begin{equation}
\frac{1}{V} \langle R^2 \rangle_N = \int {\rm d}^D X \ 
\vec{X}^2 \ G^{(2)}_N(\vec{X}) \: .  
\label{gyrr2}
\end{equation}
Since the gyration radius is a translationally invariant
quantity we have to normalize it with the total volume of the
target space and rather refer to its average density.  
The external Hausdorff dimension $D_{H}$ can then be read off
from the large $N$ behaviour~:
\begin{equation}
\sqrt{\langle R^2 \rangle_N } \sim N^{1/D_H} \: .
\label{DHdef}
\end{equation}
The dependence on the volume $V$ is hidden in an $N$-independent
constant which is not displayed in the last formula. 
The symbol $\sim$ referes to the leading behaviour.

For trees of sufficently large size $N$ one can also
define a local fractal dimension $D_L$ \cite{bpj} 
of the external geometry by measuring
the average number of vertices within a spherical 
shell of radius $X=|\vec{X}|$ from the scaling window
$X_{UV} \ll X \ll a N^{1/D_H}$
, which is defined above 
the ultraviolet cut-off scale and below the infrared scale.
As follows from the definition (\ref{gNn})
for the case of two-point function, 
the number of vertices within a spherical shell of
width ${\rm d} X$ is given by the two-point function~\cite{bpj}~:
\begin{equation}
\label{vis}
n(X) \: {\rm d} X \sim X^{D-1} \: G^{(2)}_N(X) \: {\rm d} X  
\sim X^{D_{L}-1} \: {\rm d} X\: .
\end{equation}
Here we have used the fact that the 
the two-point function is spherically symmetric, i.e.
$G^{(2)}_N(\vec{X}) = G^{(2)}_N(X)$.
The integral over the $D$-dimensional angular part is included in
the proportionality constant.  
In the large $N$ limit one expects the existence of 
a window $X_{UV} \ll X \ll a N^{1/D_H}$
where the two-point function exhibits the scaling 
behaviour~\cite{bpj}~: 
\begin{equation}
\label{tpfsb}
G^{(2)}_{N}(X) \sim X^{-\delta} \: .
\end{equation}
If $\delta$ is negative then $G^{(2)}_{N}(X)$ behaves
as a slow varying function of $X$, which can 
in a narrow range and with some corrections 
be viewed as a constant $G^{(2)}_{N}(X) \sim 1$.
Thus depending on the value of $\delta$~ we have~:
\begin{equation}
n(X) \, {\rm d} X \sim X^{D_{L}-1} \, {\rm d} X 
\sim \left\{ \begin{array}{lll} X^{D-1} \, {\rm d} X
& {\rm for} & \delta \le 0 \:  \\
                            X^{D-1-\delta} \, {\rm d} X & 
{\rm for} & \delta > 0 \: . \end{array} \right.
\label{shell}
\end{equation}
In the first case $(\delta \le 0)$ the number of 
points in a spherical shell $n(X)$ 
grows with the power of the canonical dimension $D$.
Only in the second one $(\delta > 0)$ the fractal nature
leaves traces in the calculation of $D_L$. 

\section*{Universality classes and singularity types}

In this section we will shortly summarize 
results concerning the classification of the
scaling behaviour according to the internal 
geometry of the trees \cite{net2,ft1}. 
One defines the critical exponent $\gamma$ 
of the grand-canonical susceptibility via~:
\begin{equation}
\chi_{\mu} = 
\frac{{\partial}^2 {\cal Z}}{{\partial} \mu^2} 
\sim \Delta \mu^{-\gamma} \: ,
\end{equation}
where $\Delta \mu \equiv \mu - \mu_{0}$ 
controls the behaviour of the partition
function at the radius of convergence 
$g_{0}=e^{-\mu_0}$ of the series (\ref{grandz}). 
Here $\mu_{0}$ is the critical 
value of the chemical potential. 
If $\gamma$ is positive the susceptibility
$\chi_\mu$ itself is divergent at $\mu_0$. 
If it is negative, the right-hand-side of the last
equation should be understood as the most singular 
part of the susceptibility, which after taking higher 
derivatives, will give the leading divergence. 
The primary classification of the universality 
classes for models of branched geometry is
based on the value 
of the susceptibility exponent $\gamma$.

The susceptibility exponent gives the subexponential
behaviour of the canonical coefficents 
$z_{N}$ for large $N$~:
$z_N \sim N^{\gamma-3} \exp (\mu_0 N)$. Indeed, if one 
inserts this form to the definition of the partition
function (\ref{grandz}), one obtains~:
\begin{equation}
\chi_{\mu} = \frac{{\partial}^2 {\cal Z}}{{\partial} \mu^2} =
\sum_{N} N^2 z_N e^{-\mu N} \sim 
\sum_{N} N^{\gamma-1} e^{-\Delta \mu N} \sim \Delta \mu^{-\gamma} \: .
\label{susc} 
\end{equation}

The susceptibility $\chi_{\mu}$ is proportional
to the first derivative of the 
grand-canonical partition function $\Phi$
for planted rooted trees, defined 
by equation (\ref{caphi}) in the appendix~:
$\chi_{\mu} \sim \partial \Phi/\partial \mu$. 
The reason, why this relation 
is useful is, that there exists a closed
relation -- 
a so-called self-consistency relation (\ref{sc}) 
-- for $\Phi$~:
\begin{equation}
g = g(\Phi) = \frac{\Phi}{\sum_{q=0}^\infty \frac{w_{q+1}}{q!} \Phi^q} \: ,  
\label{gofphi}
\end{equation}
which can be inverted for $\Phi = \Phi(g)$ and from 
which one can extract the singular part of $\Phi$~:
$\Phi \sim  \Delta g^{1-\gamma}$ and hence also of
$\chi_\mu$. Note that the denominator of 
(\ref{gofphi}) is nothing but the first derivative 
of the potential $V(\Phi)$ defined 
(\ref{V}) within the appendix.
 
One can invert the function $g(\Phi)$ in the region where 
it is strictly mono\-tonous. Generically $g(\Phi)$
grows monotonically from zero for $\Phi=0$ to
some critical value $g_0$ 
at $\Phi_0$ where $g'(\Phi_0)=0$. 
Clearly the inverse function $\Phi$ has a square 
root singularity at $g_0$~: $\Phi \sim \sqrt{\Delta g}$
then. It follows that $\gamma=1/2$ for the class of
generic trees. 

The region of the monotonous growths of 
the function on the right hand side 
of (\ref{gofphi}) may be limited
by $\Phi_0$ being the radius of convergence
of the series in the denominator of $g(\Phi)$.

The inverse function $\Phi(g)$ 
is then singular at $g_0=g(\Phi_0)$ 
with a singularity inherited from the 
singular behaviour of $g(\Phi)$ at $\Phi_0$. 
It can be shown that in this case 
the susceptibility exponent is negative
and the corresponding trees are collapsed.

In the marginal situation the two conditions
which limit the region of the monotonous growth
of $g(\Phi)$ work collectively at a point 
$\Phi_0$ being at the same time 
the radius of convergence of the series in the
denominator of $g(\Phi)$ and
the zero of the derivative $g'(\Phi_0)=0$.
In this case the exponent $\gamma$ can assume
any value within the interval $[0,1/2)$.
Trees which belong to this class are called
scale-free \cite{net2}. \\

The three classes correspond to different scaling
behaviours of the two-point function as will be
discussed in the following section.

\section*{Internal two-point function}

In this section we shall calculate the 
two-point function for the internal geometry. 
This function will enable us to determine 
the scaling and the fractal properties of the 
internal geometry of tree graphs. 
As we will show they are different
for generic, collapsed and scale-free trees.

In the appendix we deduce 
an explicit formula (\ref{tpdef}) 
for the grand-\-ca\-no\-ni\-cal two-point
function ${\cal G}^{(2)}(\mu,n)$. 
This function is singular for 
$\Delta \mu = \mu\! -\! \mu_0 \rightarrow 0^+$,
and singularity is related to the large $N$-behaviour
of the canonical two-point function 
${\cal G}^{(2)}_N \left(n\right)$.
The singularity of ${\cal G}^{(2)}(\mu,n)$
can be determined directly
from the identity (\ref{tpdef}) by inserting 
the most singular part of $\Delta \Phi$ to
$V'(\Phi)$ and $V''(\Phi)$. Here we will show an
alternative way using a standard scaling
argument from statistical mechanics \cite{scaling}.
Denote the singularity exponent of the two-point
funtion by $\nu$~:
\begin{equation}
{\cal G}^{(2)} \left(\mu,n \right) 
\sim \exp \left( - c (n+1) \Delta \mu^\nu \right) \: ,
\label{gsing}
\end{equation}
where $c$ is a constant which only 
depends on the particular choice of the weights $w_{q}$. 
The exponent $\nu$ is usually
called mass exponent. 
Summing over distances $n$ 
we obtain the susceptibility (\ref{susc})~:
\begin{equation}
\chi_\mu = \sum_n {\cal G}^{(2)} \left(\mu,n \right) \sim 
\int {\rm d} n \ \exp \left( - c n \Delta \mu^\nu \right) 
\sim \Delta \mu^{-\nu} \: .
\label{integ}
\end{equation}
According to the definition (\ref{susc}), the susceptibility exponent 
is $\gamma$. Thus we have 
\begin{equation}
\nu=\gamma \: .
\label{gammanu}
\end{equation}
This relation is the Fisher scaling relation for this case.
Since we have already determined $\gamma$ we do not
have to calculate $\nu$ additionally. 
The scaling argument 
given above holds only for positive $\gamma$, 
since in this case the susceptibility is divergent
and the divergent part dominates the small $\Delta \mu$
behaviour. For negative $\gamma$
the left hand side of (\ref{susc}) is not divergent 
which means that it behaves as a 
constant as $\Delta \mu$ goes to zero. 
In this case, if one compares the result
of the integration (\ref{integ}) to the leading behaviour
of (\ref{susc}), one shall effective see that $\nu=0$.
This is what happens in the collapsed phase.

Now, inserting the most singular part of the 
grand-canonical two-point function (\ref{gsing}) 
to the inverse Laplace transform (\ref{GIT}) we can deduce
the large $N$ behaviour of the canonical two-point function~:
\begin{equation}
{\cal G}^{(2)}_N\left(n\right) 
\sim e^{+\mu_{0} N} L_\nu(cn,N) \: ,
\label{GL2}
\end{equation}
where 
\begin{equation}
L_\nu(cn,N) =
\frac{1}{2\pi i} \int^{\xi_r + i\infty}_{\xi_r - i\infty}
{\rm d} \xi \ e^{-c n \xi^\nu +   \xi N}
\label{levy}
\end{equation}
is the L\'evy distribution with the index $\nu$, the 
maximal asymmetry and the range 
$C= c n \cos \left( \pi \nu/2 \right)$ \cite{levy1,levy2}.

The large $N$ asymptotic behaviour of $L_\nu(cn,N)$ with $\nu<1$
is given by the following series \cite{levy2}~:
\begin{equation}
L_\nu(cn,N) = \frac{1}{\pi N} 
\sum_{k=1}^\infty (-)^{k+1}
\left( \frac{cn}{N^\nu} \right)^{k} 
\frac{\Gamma(1+k\nu)}{\Gamma(1+k)} \sin(\pi\nu k) \: .
\label{levyseries}
\end{equation}
For large $N$ and fixed $n$, the first term
dominates the behaviour of the series~:
\begin{equation}
L_\nu(cn,N) \stackrel{N\rightarrow \infty}{\sim}
 \frac{\nu \Gamma(\nu) \sin(\pi\nu)}{\pi}
\frac{c n}{N^{1+\nu}} \, .
\label{Lmu}
\end{equation}
We see from the formulae (\ref{GL2}) and (\ref{levy}) 
that the two-point correlation function 
in the large $N$ limit is effectively a 
function of the argument $cn/N^\nu$. 
Indeed, if one changes the integration 
variable $\xi$ in (\ref{levy}) to $\xi'= \xi N$ 
one obtains~:
$L_\nu(cn,N) = N^{-1} L_\nu(cn/N^\nu,1) = N^{-1} l_\nu(u)$,
where $l_\nu(u)$ is a function of a single argument 
$u=\nu cn/N^\nu$. For later convenience we also included
$\nu$ into the definition of the universal argument.
Using the saddle point approximation
to the integral (\ref{levy})  
one can find that for large $u=\nu cn/N^\nu$ 
the function $l_\nu(u)$ leads to:
\begin{equation}
{\cal G}^{(2)}_N\left(n\right) \sim \frac{1}{N} e^{+\mu_{0} N} l_\nu(u) 
= \frac{\sqrt{a}}{\sqrt{2\pi} N} e^{+\mu_{0} N} \ 
u^{a/2} \exp \left( -b u^a \right) \, ,
\label{spoint}
\end{equation}
where $a=1/(1-\nu)$ and $b=(1-\nu)/\nu$. 
The average internal distance between two vertices can then 
be calculated~:
\begin{equation}
\label{unaverage}
\langle n \rangle_{N} = \sum_n n g^{(2)}_N(n) = 
\frac{N^\nu}{\nu c} 
\frac{\int_0^\infty {\rm d} u \ l_\nu(u) \ u }{\int_0^\infty 
{\rm d} u \ l_\nu(u)}  \, .
\end{equation}
Comparing the $N$-dependence on the right hand side
of this equation to the $N$-dependence on the
right hand side of equation (\ref{ddH}),
which defines the internal Hausdorff dimension $d_H$,
we see that $d_H$ is the inverse of $\nu$~:
\begin{equation}
\label{critexprel}
d_H = \frac{1}{\nu} = \frac{1}{\gamma} \, .
\end{equation}

Thus, the Hausdorff dimension is $d_H=2$ for generic trees.
For scale-free trees $d_H$ changes continuously from $2$ to $\infty$
since $\gamma$ belongs to the interval $\left[ 0 , 1/2 \right)$ 
then \cite{net2,bp2,ft1}.

On the other hand we see from (\ref{Lmu})
that for large $N$ and small $n$ the normalized 
two-point function grows linearly with $n$, i.e.~:
\begin{equation}
\label{g2asym}
g^{(2)}_{N\rightarrow \infty}(n) \sim n \: .
\end{equation}
The normalization coefficent behaves as $c/N$ in the
large $N$ limit. Since the sum over 
this function is proportional
to the number of vertices in the distance $n$ 
from a given vertex,
the last formula tells us that the 
local Hausdorff dimension is $d_L=2$. 
We see that locally for sufficently large $N$, 
it is difficult to distinguish 
the scale-free trees from the generic ones 
by measuring short internal distances, 
since both classes have the same internal 
Hausdorff dimension $d_L=2$.
One has to go to large
distances to see different scaling properties
depending on the type of the ensemble (\ref{spoint}).
For collapsed trees the Hausdorff dimension
is infinite. In this case, the universal scaling argument
$u$ of the two-point function is proportional to $n$ but it
does not depend on $N$. This is related to the fact
discussed before that the effective value of the exponent
$\nu$ is equal zero.

The saddle point approximation (\ref{spoint}) actually 
gives the exact result for $\nu=1/2$ for the whole
range of $u$. The reason for this is that in this case 
the integrand of the approximated expression (\ref{levy})
is gaussian. For some specific values of $\nu$
one can express the
Laplace transform (\ref{levy}) in terms of special 
functions. For example for $\nu=1/3$~:
\begin{equation}
\label{besselsol}
L_{1/3}(cn,N) = \frac{1}{N} 
\frac{\sqrt{3}}{\pi} u^{3/2} K_{1/3} \left( 2 u^{3/2} \right) \, ,
\end{equation}
where $u = \nu cn/N^\nu$.
For large $u$ the saddle point formula 
(\ref{spoint}) coincides with this one, 
while for small $u$ the two functions deviate 
a little from each other.

\section*{Gaussian trees}

Now we can determine the properties of the external 
geometry of gaussian trees. In this case,
the weight $f(\vec{x})$ in the partition 
function (\ref{zn}) for embedding 
a link $\vec{x}$ is given by a gaussian function.
The function has a vanishing mean~:
\begin{equation}
f(\vec{x}) = (2\pi \sigma^2)^{-D/2} 
\exp \left( -\frac{\vec{x}^2}{2\sigma^2}
\right) \: .
\label{f1x}
\end{equation}
In other words, for gaussian trees the link vectors 
are independent identically distributed 
gaussian random variables. 

As a consequence, the probability 
density $f_n(\vec{X})$ (\ref{convo}) for the endpoints of 
the path of length $n$ on a tree to have the relative 
coordinate $\vec{X}$ is given by~:
\begin{equation}
f_n(\vec{X}) = (2\pi n \sigma^2)^{-D/2}
\exp \left( -\frac{\vec{X}^2}{2n\sigma^2} \right) = 
\left(\frac{1}{\sqrt{n}}\right)^D \, 
f\left(\frac{\vec{X}}{\sqrt{n}}\right) \: .
\label{fnx}
\end{equation}
This follows from the stability of the gaussian distribution
with respect to the convolution. 
Inserting the function $f_n(\vec{X})$ to the formulae 
(\ref{G2}), (\ref{G3}), {\em etc.} we can determine
the multi-point correlation functions for gaussian trees.
In particular, if we insert (\ref{fnx})  
to (\ref{G2}), we obtain in the large $N$ limit~:
\begin{equation}
G^{(2)}_N(\vec{X}) = \frac{c}{N(2 \pi \sigma^2)^{D/2}}
\sum_{n=1}^\infty \: 
n^{1-D/2} \exp \left(-\frac{\vec{X}^2}{2n\sigma^2} 
-\frac{c n^2}{2N} \right) \: .
\label{GG2}
\end{equation}
Here we used the same approximation for the 
internal two-point function as in the discussion of 
equation (\ref{ggn2}) in the appendix. 
This is a good approximation
for large $N$. Additionally, we substituted the 
upper limit $N-1$ of the summation over $n$ by 
infinity. This introduces small corrections 
which disappear exponentially in the large $N$ limit.

In order to measure the external Hausdorff dimension
$D_H$ we have to determine the dependence of the
expectation value $\langle R^2 \rangle_N$ of the gyration 
radius on the system size $N$. The expectation value can
be calculated by integrating the two-point function
over $\vec{X}$ as in equation (\ref{gyrr2}).
If one first integrates over $\vec{X}$ before 
summing over $n$ one obtains~:
\begin{equation}
\label{Ggyrr1}
\frac{1}{V} \langle R^2 \rangle_N = 
\frac{c D \sigma^2}{N} \sum_{n=1}^\infty n^2 
\exp \left( -\frac{c n^2}{2N} \right)\: .
\end{equation}
One can approximate the right-hand side by replacing 
the summation from $1$ to $\infty$ through an 
integration over the whole positive real axis~:
\begin{equation}
\label{Ggyrr2}
\frac{1}{V} \langle R^2 \rangle_N = 
\frac{c D \sigma^2}{N} 
\int_0^\infty {\rm d} y \ y^2 \exp \left(-\frac{cy^2}{2N} \right) 
\sim D \sigma^2 \sqrt{\frac{N}{c}} \: . 
\end{equation}
We see that the typical extent of the
system $\sqrt{\langle R^2 \rangle_N}$ grows as $N^{1/4}$
and hence the Hausdorff dimension for 
generic gaussian trees is $D_H=4$.

More generally, in order to determine the
dependence of the expectation value of the 
gyration radius on the system size for any type of trees 
one can first calculate the second moment of 
the function $f_n(\vec{X})$~:
\begin{equation}
\label{X2fnaverage}
\langle X^2 \rangle_n \equiv \int {\rm d}^D X \ \vec{X}^2 f_n(\vec{X}) \: ,
\end{equation}
which corresponds to the average extent of the path built 
out of $n$ links of the tree. 
The insertion of this result to (\ref{gyrr2}) yields~:
\begin{equation}
\label{gyrrvsX2average}
\frac{1}{V}
\langle R^2 \rangle_N = 
\sum_{n} \langle X^2 \rangle_n \ g_N^{(2)}(n) \: .
\end{equation}
Since for gaussian weights the second moment
$\langle X^2 \rangle_n$ is proportional to $n$, 
i.e. $\langle X^2 \rangle_n \sim n$, the following
relation holds \cite{foot2}~:
\begin{equation}
\label{GDHdef}
\langle R^2 \rangle_N \sim \sum_{n} n g_N^{(2)}(n) = 
\langle n \rangle_{N} \sim N^{1/d_H} \: ,
\end{equation}
from which we conclude that the external 
and internal Hausdorff dimensions are
related by~:
\begin{equation}
\label{GDHvsdH}
D_H = 2 d_H 
\end{equation}
for gaussian trees. This relation holds for generic, 
scale-free and collapsed trees.
This for example means that the Hausdorff dimension 
$D_{H}$ of collapsed trees is infinite, or in other words, 
that the target space extent of the system 
does not change with the number of vertices on the tree.

Let us come back to generic gaussian trees. We will
calculate the local Hausdorff dimension $D_L$ and
compare it with $D_H=4$. The starting point of this
calculation is equation (\ref{shell})
which relates the number of points within 
the shell of radius between $X$ and $X+{\rm d}X$  
to the behaviour of the two-point function 
in the scaling window $X_{UV} \ll X \ll X_{IR}$,
where $X_{UV} \sim \sigma$, $X_{IR} \sim a N^{1/4}$.

The two-point function (\ref{GG2}) is a decreasing
function. It has a cut-off at $X \sim a N^{1/4}$ as
follows from the scaling arguments.
The large $n$-part of the sum (\ref{GG2})
over $n$~: $n \gg 1$ can be approximated by 
an integral over $n$. This part of the sum
has a significant contribution if $\zeta=X^2/\sigma^2 \gg 1$.
Thus for $X \gg \sigma$ the dominant dependence 
of the sum on $X$ can be approximated by~:
\begin{equation}
G^{(2)}_{N}(X) \sim 
\int_{c_1}^{c_2\sqrt{N}} {\rm d} n 
\ n^{1-D/2} e^{-\zeta/n}  \sim \zeta^{2-D/2} \: .
\end{equation}
with some constants $c_1, c_2$.
The upper limit of the integral 
comes from the term $\exp(-c n^2/2N)$.
The exact shape of the integrand at 
large $n \sim \sqrt{N}$ is
unimportant for $D>4$, 
because the dominating behaviour
$G^{(2)}_{N}(X) \sim \zeta^{2-D/2} \sim X^{4-D}$
is due to the lower limit of the integration.
This scaling form of $G^{(2)}_{N}(X)$ 
breaks down for short distances $X$ of order 
$\sigma$ and for large 
$X$ of the order of 
the infrared cut-off $aN^{1/4}$. 
When $X$ is of the order $\sigma$
the integrand is a sum of gaussians 
of widths larger than 
$X$ and hence $G^{(2)}_{N}(X)$ 
is a slow varying function.

\begin{figure}[htbp]
\centerline {\epsfig {file=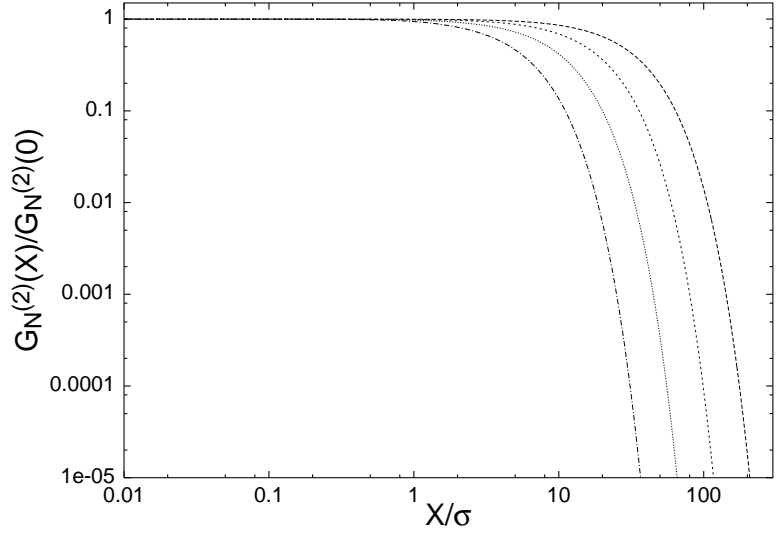, 
bbllx=40, bblly=40, bburx=410, bbury=302, width=10.5cm, angle=0}}
\caption[d2fig]{
The normalized two-point function
for $D=2$ and for $N=10^3,10^4,10^5$ and $10^6$, from left to right,
respectively. 
The functions are constant in the region of small $X$.
This region extends to some cut-off whose position 
grows with the power $N^{1/4}$.}
\label{d2fig}
\end{figure}

\begin{figure}[htbp]
\centerline {\epsfig {file=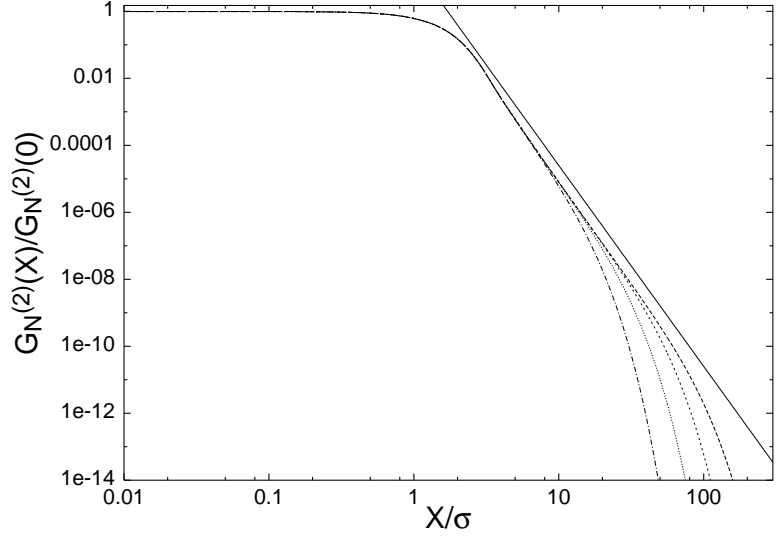, 
bbllx=40, bblly=40, bburx=410, bbury=302, width=10.5cm, angle=0}}
\caption[d2fig]{
The normalized two-point function
for $D=10$ and for $N=10^3,10^4,10^5$ and $10^6$, from left to
right, respectively. 
The functions are constant in the region of small $X$.
Then they develop a scaling part (linear in the figure)
in which they behave as $\propto X^{-6}$.  
For a comparison of the slopes we also 
displayed (solid line) a pure power-law $\propto X^{-6}$.}
\label{d10fig}
\end{figure}

For $D \le 4$ the regime changes. The divergence at small $n$
disappears and the terms for large $n$,
$n \sim \sqrt{N}$, dominate in the sum. 
The sum (\ref{GG2}), viewed as a function 
of $X$, looks as a sum of gaussians 
whose arguments $X$ are maximally of 
the order of the widths.
This is a slow varying function of $X$ 
for $X^2/\sigma^2 \sqrt{N}$. Hence we expect that 
as long as $X \ll N^{1/4}$ it is almost constant 
$G^{(2)}_{N}(X) \sim 1$. As an example we performed
the sum (\ref{GG2}) numerically for $N$ up 
to $10^6$.
The results presented in the figures \ref{d2fig} and 
\ref{d10fig} corroborate the anticipated 
behaviour of $G^{(2)}_{N}(X)$ by the 
arguments given above
 \cite{foot3}.

As a consequence we see that
the number of points within the spherical shell
(\ref{shell}) depends on the radius $X$ as~:
\begin{equation}
\label{Gextss}
n(X) \sim \left\{ \begin{array}{lll} X^{D-1} & {\rm  for} & D \le 4 \:  \\
                            X^3 & {\rm  for} & D>4 \: . \end{array} \right.
\end{equation}
This leads to the following result for the Hausdorff dimension~:
\begin{equation}
\label{GDL}
D_L
= \left\{ \begin{array}{lll} D & {\rm  for} & D \le 4 \: , \\
                           4 & {\rm  for} & D>4 \: . \end{array} \right.
\end{equation}
In other words, the fractal dimension $D_L$ 
measured by the local observer, 
is equal to the global one, 
i.e. $D_H=D_L$, if the dimension
of the target space is large enough. 
If the target space dimensionality
is too small, the fractal structure cannot develop.
One can understand this in the following way. For
trees embedded in a $D<4$ dimensional target space, 
vertices of the tree lie in a ball with a radius 
proportional to $N^{1/4}$. 
There are $N$ vertices within the ball 
while the volume of the ball is proportional 
to $N^{D/4}$. This means that for large $N$, 
the vertices deep inside the ball are densily 
and uniformly packed. A local observer who surveys
only a small region far from the ball boundary
will see uniformly distributed vertices in a $D$ dimensional 
space. As a consequence he or she will measure $D_L=D$. 
The situation changes for $D>4$, 
because then the volume of the ball is
proportional to $N^{D/4}$ and hence grows much faster 
than the the number of the vertices. 
In the large $N$ limit the volume of the ball
will therefore be large enough to let the system
develop a loose fractal structure. 

Similarly, we expect that for the scale-free trees,
the local Hausdorff dimension is 
\begin{equation}
\label{sfDL}
D_L
= \left\{ \begin{array}{lll} D & {\rm  for} & D \le D_H \: , \\
                           D_H & {\rm  for} & D>D_H \: . \end{array} \right.
\end{equation}

Since the Hausdorff dimension $D_{H}$ is 
infinite for collapsed trees, 
one cannot define a local 
Hausdorff dimension $D_L$ in the 
same manner as above, because the 
infrared cut-off is a constant in this 
case.

\section*{L\'{e}vy trees}
So far we have only considered gaussian embedding weights 
$f(\vec{x})$ (\ref{f1x}). In this case links typically
have the length $\sigma$ and one can hardly find a link 
on a tree longer than $3\sigma$. In other words the energy 
costs for the embedding of long links are so high 
that those links do not appear. 
One can, however, consider models with weights 
which allow long links. In the next section
we will discuss this issue in a more general context,  
while in this section, as toy models, we will consider models 
of trees embedded in $D=1$ dimensional target space 
with the weigths given by 
a symmetric L\'evy distribution \cite{levy1,levy2}. 
Despite their simplicity, the models 
with those weights already basically capture all 
interesting features of more complicated models. 
The weights read~:
\begin{equation}
f(x) = L_\alpha(A,x) = 
(2\pi)^{-1} \int^{+\infty}_{-\infty} 
{\rm d} p \ \exp \left( -A^\alpha |p|^\alpha - i p x \right) \: ,
\label{flevy}
\end{equation}
with $\alpha$ from the interval $(0,2]$. In the limiting
case $\alpha=2$ is a gaussian distribution with
the width $\sigma = \sqrt{2} A$, 
and for $\alpha=1$ is a Cauchy distribution. 

The weights are symmetric stable distributions with
the stability index $\alpha$.
Here we are interested only in symmetric functions
$f(x)=f(-x)$ because the links are unoriented.
This implies that any function 
defined on them has
the property~: $f(r_{ij}) = f(-r_{ij})=f(-r_{ji})=f(r_{ji})$.

The distribution (\ref{flevy}) is stable with respect to
the convolution~:
\begin{equation}
\label{convolute}
L_\alpha(A,x) = \int^{+\infty}_{-\infty} {\rm d} x_1 \ 
L_\alpha(A_1,x_1) L_\alpha(A_2,x-x_1) \: ,  
\end{equation}
where $A^\alpha = A_1^\alpha + A_2^\alpha$.
If one repeats this for the convolution of $n$ identical terms
to calculate $f_n(X)$ (\ref{charact}) 
one can see that the function $f_n(X)$ is given by 
a rescaled version
of the function for a single link $f(X)$~:
\begin{equation}
f_n(X) = \frac{1}{n^{1/\alpha}} f\left(\frac{X}{n^{1/\alpha}}\right) \: . 
\label{ffn}
\end{equation}
Now we can combine this scaling with the scaling of the
internal two-point function which, as we know, is a function 
$N^{-1} l_\nu(v)$ of a scaling variable $v = n/N^\nu$,
to deduce the scaling of the external two-point 
function (\ref{G2})~:
\begin{equation}
\label{LG2scal}
G^{(2)}_N(X) = \sum_n f_n(X) g^{(2)}_N(n) \sim
\frac{1}{N} \int \frac{{\rm d} n}{ n^{1/\alpha}}
f\left(\frac{X}{n^{1/\alpha}}\right) 
l_\nu\left(\frac{n}{N^\nu}\right) \: .
\end{equation}
The result of the integration can be written 
as a function of an argument $X/N^{\nu/\alpha}$
with some prefactor depending on $N$.
As a consequence one expects the external 
Hausdorff dimension to be~:
\begin{equation}
\label{LDHvsdH}
D_H = \alpha/\nu = \alpha d_H \, .
\end{equation}
The case $\alpha=2$ was discussed before. Despite similarities,
the case $\alpha<2$ is different from the gaussian one,
since in this case the distribution
$f(x)$ has a fat tail for large $x$~:
\begin{equation}
{\rm d} x \ f(x) \sim \frac{{\rm d} x}{x} 
\frac{A^\alpha}{x^\alpha} \: ,
\label{tail}
\end{equation}
which according to the scaling (\ref{ffn}) is equally
important in $f_n(X)$ for any $n$ .
The second moment $\langle X^2 \rangle_n$
of the distribution $f_n(X)$ is infinite. 
As a consequence, also the gyration radius is infinite. 
One has to find an alternative measure
of the linear system extent in order to define the Hausdorff 
dimension $D_H$. A natural candidate for such a quantity 
is~:
\begin{equation}
\label{Rq}
R^q = \frac{1}{N^2} \sum_{i,j} |X_i - X_j|^q,
\end{equation}
for $q<\alpha$. The Hausdorff dimension can
now be calculated from the large $N$ behaviour of this
quantity~:
\begin{equation}
\label{LDHdefRq}
\langle R^q \rangle_{N} \sim N^{q/D_H} \: .
\end{equation}
Using the same arguments as for the gaussian case,
one can check that the following relations hold
\begin{equation}
\frac{1}{V} \langle R^q \rangle_N = \int {\rm d} X \ 
|X|^q \ G^{(2)}_N(X) = 
\sum_n \langle |X|^q \rangle_n \ g^{(2)}_N(n) \: ,
\label{gyrrq}
\end{equation}
where 
\begin{equation}
\label{Xqfn}
\langle |X|^q \rangle_n = 
\int {\rm d} X \ |X|^q f_n(X) \sim n^{\frac{q}{\alpha}} \, .
\end{equation}
It follows, that 
$\langle R^q \rangle_N \sim N^{q/{\alpha d_H}}$
and hence $D_H = \alpha d_H$, as already mentioned.

\section*{Other trees}

We will continue the discussion of the one-dimensional case, i.e. $D=1$.
The extension to higher dimensions $D$ shall afterwards
be straightforward. 
The embedding weights for links may in general 
be given by any normalizable non-negative 
symmetric function~: 
$f(x)=f(-x)\ge 0$ such that~: $\int {\rm d} x \ f(x) =1$. 

We are interested in the emergence of the scaling properties
for large $N$. From the considerations of the internal
geometry we know that the internal distance between
two random vertices on the tree, $n \sim N^{1/d_H}$, 
grows with $N$ unless the trees are collapsed. 

We also know that between those 
two random vertices we can
draw a unique path on the tree. This path can be
treated as a random path of $n$ links. 
So, in a sense, we are interested in the probability distribution 
that the remote ends of the random path with $n$ links
have the the relative position $X$ in the embedding space.
In particular we are interested in 
the limit $n\rightarrow \infty$.
This probability distribution is given by $f_n(X)$. 
For large $n$ the function $f_n(X)$ can be determined from the 
central limit theorem. Roughly speaking, if the second moment
of $f(x)$ exists, $f_n(X)$ approaches a gaussian 
distribution with the variance $\sigma_n = \sqrt{n}\sigma$,
otherwise $f_n(X)$ approaches
the L\'evy distribution (\ref{flevy}) with the
scale parameter $A_n = n^{1/\alpha} A$.
Thus, if a distribution has a power-tail
$f(x) \sim x^{-1-\alpha}$ for large $x$, 
the limiting distribution $f_n(X)$ for
large $n$ will approach the gaussian distribution
if $\alpha>2$, or the L\'evy distribution (\ref{flevy})
if $\alpha<2$ \cite{levy1,levy2}.
The limiting case $f(x) \sim x^{-3}$ belongs
to the gaussian domain of attraction but it has a logarithmic
anomaly of the variance which in this case
does not grow as $\sqrt{n}$ but faster, i.e.
with some additional logarithmical factor of $n$.

For $\alpha > 2$ the approach of $f_n(X)$ 
to the gaussian distribution for large $n$ 
is non-uniform and takes place in the 
central region of the distribution $f_n(X)$ for
$|X| < X_*$, where $X_*$ scales with $n$ as
\begin{equation}
\label{borderscal}
X_* \sim b \sqrt{n \log{n}} \: .
\end{equation}
Here $b$ is some constant \cite{foot4} representing
a scale of the distribution.
Outside this region $f_n(X)$ deviates severe from
the normal law, and in particular it preserves the
power-law tail for $X \gg X_*$~:
\begin{equation}
{\rm d} X \ f_n(X) \sim \frac{{\rm d} X}{X} 
\frac{n A^\alpha}{X^\alpha} \: ,
\label{ntail}
\end{equation}
with a tail amplitude proportional to $n$. In other
words, for any finite $n$ the power-law tail 
is present in the distribution $f_n(X)$.
Therefore all absolute moments 
of order $Q > \alpha$ of this distribution
are infinite~:
\begin{equation}
\label{XQdiv}
\langle |X|^Q \rangle_n = \infty \, 
\end{equation}
for finite $n$, and as a consequence also
\begin{equation}
\label{RQdiv}
\langle R^Q \rangle_{N} 
= \int {\rm d} X \ X^Q G^{(2)}_N(X) 
 = \sum_n \langle |X|^Q \rangle_n g^{(2)}_N(n) 
= \infty \, .
\end{equation}
For $N \rightarrow \infty$ the sum is dominated
by terms of large $n \sim N^{1/D_H}$. 
For $n \rightarrow \infty$ the distribution 
$f_n(X)$ becomes normal in the whole region 
from $-\infty$ to $+\infty$.
Indeed, the ends of the central region $X_*$
move to infinity faster than the variance $\sigma_n \sim \sqrt{n}$
and the contribution coming from the outside of 
the central region $|X| > |X_*|$ disappears as~:
\begin{equation}
\label{corrtogauss}
\int_{|X|>X_*} {\rm d} X \ f_n(X) 
\sim \frac{1}{n^{\alpha/2-1} \log^{\alpha/2} n} \, .
\end{equation}
Thus the non-gaussian part
including the tails becomes marginal and can be
neglected. What is left over for $n=\infty$ 
is a gaussian distribution with all moments defined. For
example, even integer moments $Q=2K$ are~:
$\langle X^{2K} \rangle_n = (2K-1)!! \ (n \sigma^2)^K$.
Thus, after taking this limit the trees behave like 
gaussian ones. This limit is subtle, because as long as 
$N$ (and $n$) is large but finite  higher moments,
$Q > \alpha$, of the distribution are infinite. 

Now we shall shortly discuss the model 
in $D>1$ dimensions.
As before we consider spherically symmetric distributions
$f(\vec{x}) = f(x)$, 
$x= |\vec{x}|$ which have
a power-law dependence for large lengths of the link
vectors $f(x) \sim x^{-D-\alpha}$~:
\begin{equation}
{\rm d}^D x \ f(\vec{x}) = 
{\rm d} \Omega_D \ {\rm d} x \ x^{D-1} f(x) \sim
{\rm d} \Omega_D {\rm d} x \ x^{-1 - \alpha} \, ,
\label{Dtail}
\end{equation}
where ${\rm d} \Omega_D$ is the angular part of the measure.
The main difference to the one dimensional case is that
the effective power of the link length 
distribution changes by $D-1$ due 
to the angular measure $x^{D-1}$.
Otherwise, the dependence of the scaling on $\alpha$
goes in parallel to the one dimensional case that is
the distribution belongs to the gaussian domain of attraction
if $\alpha \ge 2$ and to the L\'evy one if $\alpha < 2$. 
The characteristic function (\ref{charact})
of the corresponding limiting distribution 
is spherically symmetric~:
$\exp (-A^{\widehat{\alpha}} p^{\widehat{\alpha}})$,
where $p = |\vec{p}|$ and 
$\widehat{\alpha}={\rm min}\{2,\alpha\}$.
For example, a distribution which has a power-law tail 
$f(x) \sim x^{-1 - D}$ belongs to the domain of attraction 
of the Cauchy distribution~:
\begin{equation}
\label{cauchycase}
f(\vec{x}) = \frac{1}{(2\pi)^D}
\int {\rm d}^D q \ e^{-A \sqrt{{\vec{q}}^2} - i\vec{q}\vec{x} }
= \frac{\Gamma \left( \frac{D+1}{2} \right)}{\pi^{\left(D+1 \right)/2}} 
\frac{A}{\left(A^2 + \vec{x}^2\right)^{\frac{D+1}{2}}} \: .
\end{equation}
A new effect which arises in higher dimensions
is a possibility of a spontaneous breaking of the 
rotational symmetry. The limiting distributions for
$N\rightarrow \infty$ and the two-point function 
$G^{(2)}_N(\vec{X})$ are spherically symmetric but the 
configurations which contribute to them are not.
The effect is strong when $\alpha \le 1$
and is known from the considerations of L\'evy paths \cite{levy1}.
If we have such an ensemble of $N$ links, one can find a
link whose length is $N^{1/\alpha}$ larger 
than the sum of the lengths of the remaining links. 
This link makes the system look like a one dimensional system,
since the extent of the system in the direction of 
this link is significantly larger than in the 
other directions.
The effect becomes weaker when $\alpha$ is larger
than one. Actually it is then seen for configurations which 
come from the large $X$-tail of the two-point 
function $G^{(2)}_N(X)$. 
This configurations become marginal
for $\alpha>2$ in the large $N$ limit. 
However, as long as $N$ is finite,
the probability of large $X$ in $G^{(2)}_N(X)$ is finite
and it strongly influences the measurements of higher
moments $\langle R^Q \rangle_{N}$ of the system extent.
In other words the higher $Q$ is, 
the stronger is the contribution from the 
large $X$ part of the two-point function 
and the more contribute systems, which are elongated,
to this quantity.
In the limiting case~: $Q \rightarrow \alpha$
the main contribution to the moments $\langle R^Q\rangle_{N}$
comes from one-dimensional configurations. 

As mentioned, the branched polymer model with power-law weight
arises as the one-loop approximation of the reduced
super-symmetric Yang-Mills model \cite{ikkt2}. In particular
for $D=4$ dimensions, the embedding weights $f(x)$ behave as
$f(x) \sim x^{-D-\alpha} \sim x^{-6}$ for large link lengths
$x$. This is the marginal case $\alpha=2$ which
belongs to the gaussian domain of attraction. This means
in particular that if one first takes the limit 
$N\rightarrow \infty$ then the gaussian branched polymers emerge
for which all correlation functions are well defined. On the other
hand if one determines higher correlation functions
before one takes $N\rightarrow \infty$ one shall see 
they are divergent.

In numerical simulations of the full matrix model one
also observes power-law tails in the distribution
of the system extent and one dimensional configurations 
\cite{ssb1}-\cite{ssbx}. It is possible \cite{ssb3},
but not yet answered, that there also exists a Gaussian
limit at large $N$ in this model.

\section*{Discussion}
We investigated the model of trees embedded freely
in a $D$ dimensional target space. We classified the
scaling properties of the model by determining the
fractal dimensions for internal and external 
geometry for the ensembles of generic and exotic trees
including those which have fat tails in the distributions
of branching orders and of link lenghts in the embedding
space. We showed that, for freely embedded trees, internal
geometry is indepedent of the embedding as a result 
of the factorization (\ref{Gfg}). On the other
hand external geometry strongly depends on internal one~:
in particular the Hausdorff dimension for 
external geometry is proportional to that for internal
one $D_H=\alpha d_H$. The proportionality coefficent is given by the
stability index of the embedding weights. For gaussian
trees, in particular, it is equal two. We pointed out 
that the finite effects related to the presence of fat-tails 
lead to singularities of higher order correlation functions 
before the inifinite large $N$-limit is taken. This is a
similar effect to the one observed in the IKKT matrix 
model.

The branched polymer model captures many features of the more
complicated models of random geometry.
Despite its simplicity the model 
has a rich phase structure~: 
a generic phase of gaussian trees 
which have the Hausdorff dimensions $D_H=4$,
the phase of short trees with $D_H>4$ coming
from the embedding of scale-free and crumpled tree
graphs, and the the phase with elongated 
L\'evy branches, which has the Hausdorff dimension
$D_H<4$. 

Due to the simplicity, and the full control of the
free case, the model of branched polymers which we discussed
here, provides a good starting point for modelling
effects of non-trivial embedding like those related to
excluded volume effects or external curvature.
Such effects violate the factorization 
introducing correlations between
the internal and external geometry.
A sort of back-coupling occurs. 
The external geometry affects the 
internal one, which modifies and 
influences back the external one. 
For example, self-avoidance 
disfavours crumpled trees and hence changes
the internal Hausdorff dimension
which in turn will change the external one.

\bigskip

\par
{\bf Acknowledgements}:  
We thank P. Bialas for discussions.
This work was partially supported by the
EC IHP grant HPRN-CT-1999-000161 and by
the project 2 P03B 096 22 of the Polish
Research Foundation (KBN) for 2002-2004.
M.W. thanks the University of Bielefeld for
a graduate scholarship. Z.B. thanks the Alexander 
von Humboldt Foundation for a follow-up fellowship. 

\section*{Appendix}

This appendix summarizes 
important relations of the  
internal geometry of branched polymer 
models \cite{poly1,poly2,ft1,cenum}. 
It is intended to make this
article more self-contained.
As we already mentioned, 
the part of the model related
to the internal geometry decouples 
from the problem of the embedding 
and can hence be solved independently.
It is convenient to introduce 
several generating functions
to ease the calculations. 
Although many of the considerations 
made here are well-known, we deduce 
several relations for the generating 
functions which play important roles
in the description of branched polymer 
models. Graphical representations
of the generating functions turn out to be 
effective tools for the 
just mentioned deductions.
Finally we will be able to calculate the partition
function $z_N$ (\ref{izn}) and the two-point function 
$g^{(2)}_{N}(n)$ (\ref{g2}) of the internal geometry. 
Note that we already discussed the scaling properties 
and the universality resulting from those calculations
in the previous sections.

Recall that the internal properties of a branched polymer 
model (\ref{zn}) only depend on the internal weight 
function $W_{T}$.
For internal weights of the form (\ref{weightdef})
these properties are entirely determined when the whole 
set of branching weights $\left\{ w_q \right\}$, obeying the 
conditions (\ref{wcond}), is given.
The information about the entire set $\{ w_q \}$ of 
branching weights can alternatively be encoded in a
single function of one real variable, namely (\ref{V}).
\begin{equation}
V(\Phi) = \sum_{q=1}^\infty \frac{w_q}{q!} \Phi^q \: .
\label{V}
\end{equation}
As we shall show below, the scaling properties of the
internal geometry are directly related to the analytic
properties of this function. 
We will often refer to $V$ as a potential, since the 
most important generating functions can be
written as derivates of $V$.  

In the first section we defined 
the generating function for the 
canonical partition functions $z_N$ to be~:
\begin{equation}
{\cal Z} = \sum_{N=2}^{\infty} z_{N} g^{N} 
= \sum_{N=2}^{\infty} z_{N} e^{- \mu N} \: ,
\end{equation}
which is nothing else but the grand-canonical partition 
function for the ensemble of trees with unrestricted size. 
One reason why it is convenient 
to introduce the generating function
${\cal Z}$ is that one can write a closed self-consistency
equation for a first derivative of it, as we shall discuss below.
The grand canonical partition function can be written as~:
\begin{equation}
\label{altgrandz}
{\cal Z} = \sum_{N=2}^{\infty} \frac{1}{N!} 
\sum_{T \in {\cal T}_{N}} \left( g w_{1} \right)^{N_{1}}
\left( g w_{2} \right)^{N_{2}}\cdots 
\left( g w_{N-1} \right)^{N_{N-1}} \: ,
\end{equation}
where $N_{q}$ denotes the number of vertices of order $q$ and the 
sum begins with $N=2$ being the number of vertices on the smallest
tree.
Note that each vertex introduces a factor $\left(g w_{q}\right)$ to the
(internal) weight of the tree in the grand-canonical ensemble. 
There are two derivatives of the generating function 
${\cal Z}$ which will be useful~:
\begin{eqnarray}
\label{z1} &{\cal Z}^{(1)}& \equiv g \frac{\partial
{\cal Z}}{\partial g} = \sum_{N=2}^{\infty} N z_{N} g^{N} \: , \\
\label{caphi}
 &\Phi & \equiv \frac{1}{g} \frac{\partial {\cal Z}}{\partial w_{1}} = 
\sum_{N=1}^{\infty} \varphi_{N} g^{N} \: , 
\end{eqnarray}
where
\begin{equation}
\label{canphi}
\varphi_{N} = \frac{\partial z_{N\!+\!1}}{\partial w_1} = 
\frac{1}{\left( N\!+\!1 \right)!} 
\sum_{T \in {\cal T}_{N\!+\!1}} \frac{N_{1} W_{T}}{w_{1}} \: .
\end{equation}
Clearly, ${\cal Z}^{(1)}$ is a generating function for the canonical 
partition functions $z^{(1)}_{N} \equiv N z_{N}$ of trees with $N$ 
vertices that have 
one marked vertex. Intuitively the factor $N$ in the sum 
can be viewed as a factor which counts the possible 
choices of marking one vertex on a tree with $N$ vertices.
The derivative $\Phi$ is a generating function 
for the partition functions 
$\varphi_{N}$ of branched polymers of size $N$ having one (not
counted) additional marked vertex of order one. 
We will refer to those 
uncounted vertices as external lines. 
Because we do not count
the empty lines in (\ref{caphi}) the sum starts with $N=1$.

\begin{figure}[htbp]
\centerline {\epsfig {file=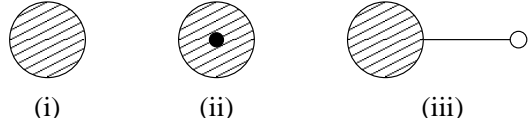, 
bburx=212, bbury=48, bbllx=0, bblly=0, width=7.0cm, angle=0}}
\caption[Graphical representations of the generating functions.]
{Graphical representations of the generating functions (a) ${\cal Z}$,  
(b) ${\cal Z}^{(1)}$ and (c) $\Phi$.}
\label{gfp}
\end{figure}

We now introduce a graphical notation for the 
generating functions with 
the following conventions~: Whenever a vertex
is represented by an empty circle, this means that this 
vertex neither introduces a weight $w_{q}$ corresponding to its order
nor a fugacity $g$. Consequently those vertices are not counted.
Solid circles correspond to counted vertices and therefore introduce 
factors $w_{q}$ and $g$.  Combinatorial factors which are due to certain
symmetries of the represented object will always be displayed explicitly. 
Links between vertices will be represented by solid lines.
As can be seen in figure \ref{gfp} the generating function ${\cal Z}$ will
be represented by a bubble, its derivative ${\cal Z}^{(1)}$
by a bubble with a filled circle inside and its derivative $\Phi$ by a bubble
with a tail having an empty circle at the end. The tail corresponds to 
the external line of the tree.

The marked vertex of the ensemble is often called root. 
Trees generated by ${\cal Z}^{(1)}$ are called rooted, 
those generated by $\Phi$ planted rooted or simply planted.

In a similar way one can also define higher derivatives 
of ${\cal Z}$.  Each derivative $g {\partial}/{\partial g}$ introduces
a new marked vertex and hence another filled circle in the bubble 
of the graphical representation. Each
derivative $\left( 1/g \right) {\partial}/{\partial w_1}$ introduces
 a new external line with an empty circle at the end. 
One could also define
derivatives $ \left( 1/g \right) {\partial}/{\partial} w_k$ 
which introduce an external uncounted vertex 
connected to the bubble via $k$ links.

The most fundamental object among all these generating functions
is the generating function $\Phi$ for planted trees.
With its help one can construct all the others. For example, 
the combination $g w_{q}\Phi^q / q!$ is a generating function 
for trees with one marked vertex 
of the order $q$. If one sums over $q$ one 
obtains a generating function for trees which have
just one marked vertex of any order. This is nothing
else but ${\cal Z}^{(1)}$ itself. Thus we have~:
\begin{equation}
{\cal Z}^{(1)} = g \sum_{q=1}^{\infty} \frac{w_{q}}{q !} \Phi^{q} 
=g \: V\left( \Phi \right) \: .
\label{nsc}
\end{equation}
If one adds a line with an empty end 
to this marked vertex 
one obtains the generating function
for planted trees. The order of the marked vertex to which 
the line is added consequently 
increases by one, i.e. $q \rightarrow q+1$.
Thus the corresponding contribution to the sum over $q$
is $g w_{q+1}\Phi^q/q!$~:
\begin{equation}
\Phi = g \: \sum_{q=0}^{\infty} \frac{w_{q+1}}{q!} \Phi^{q}  = 
g \: V' \left( \Phi \right) \: ,
\label{sc}
\end{equation}
which is a self-consistency equation for $\Phi$ from
which $\Phi$ can be calculated. Having calculated
$\Phi$ one can insert it to (\ref{nsc}) and determine ${\cal Z}^{(1)}$
and so on. 

\begin{figure}[htbp]
\centerline{\epsfig {file=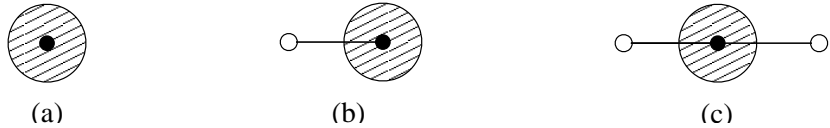, 
bburx=324, bbury=49, bbllx=0, bblly=0, width=11.0cm, angle=0}}
\caption[Graphical representation of $g V(\Phi)$, $g V'(\Phi)$, $g V''(\Phi)$]
{Graphical representation of $g V(\Phi)$, $g V'(\Phi)$ and $g V''(\Phi)$.} 
\label{phifig}
\end{figure}  

As we have seen above $g V(\Phi)$ generates trees
with one marked vertex, $g V'(\Phi)$ trees with one 
marked vertex which is connected to an external line.
One can easily check that the  $k$-th derivate 
of $V(\Phi)$~:
\begin{equation}
\label{kderV}
g V^{(k)}(\Phi) = \sum_{q=0}^{\infty} \frac{w_{q+k}}{q!} \Phi^q
\end{equation}
generates an ensemble of trees with one marked vertex
connected to $k$ external lines. 
The graphical representations of $g V^{(k)}(\Phi)$
for $k=0,1,2$ is depicted in figure \ref{phifig}.

\begin{figure}[htbp]
\centerline{\epsfig {file=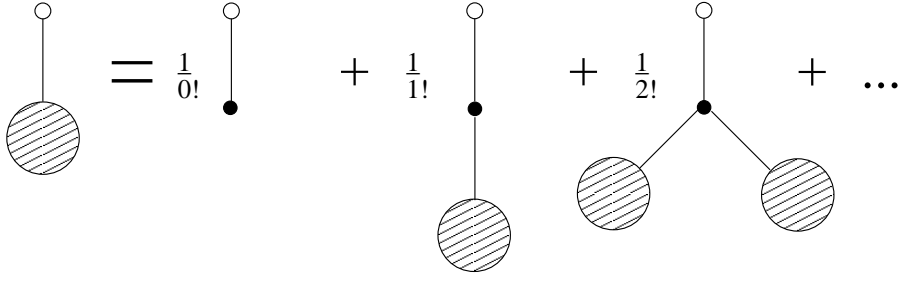, bburx=380, bbury=108, bbllx=0, bblly=0, width=12.0cm, angle=0}}
\caption[Graphical representation of the self-consistency equation.]{Graphical representation of the self-consistency equation (\ref{sc}).}
\label{scg}
\end{figure}

The self-consistency equation (\ref{sc}) is illustrated
in figure \ref{scg}. The content of
equation (\ref{nsc}) emerges automatically from
the comparison of 
figure \ref{gfp} (b) and figure \ref{phifig} (a).

Let us now illustrate how the generating function machinery works
solving the classical problem of the tree diagram 
enumeration \cite{cenum}.
We shall calculate $z_N$ for the case $W_T=1$.
This is called the Cayley problem. The number of all 
labeled trees with $N$ vertices is given by $z_N N!$. 
The self-consistency equation reduces to~:
\begin{equation}
\label{trivtreesc}
\Phi = g \: e^{\Phi} \, .
\end{equation}
This equation can be solved for $\Phi$~:
\begin{equation}
\label{trivtreephisol}
\Phi = \sum_{N=1}^\infty \frac{N^{N-1}}{N!} g^N \: .
\end{equation}
For the weights $w_q=1$ equation (\ref{nsc}) 
leads to a simple relation between ${\cal Z}^{(1)}$
and $\Phi$~:
\begin{equation}
\label{trivtreeZ1sol}
{\cal Z}^{(1)} = \Phi - g =
\sum_{N=2}^\infty \frac{N^{N-1}}{N!} g^N = 
\sum_{N=2}^\infty N z_N g^N \: .
\end{equation} From this relation 
one can calculate the 
canonical partition function~:
\begin{equation}
\label{trivtreeznsol}
z_N = \frac{N^{N-2}}{N!} \quad \mbox{for} \quad N \geq 2 
\end{equation}
and the number of labeled tree diagrams to be $N^{N-2}$.
For large $N$ one can approximate $z_N$ by~:
\begin{equation}
z_N = \frac{N^{N-2}}{N!} \sim (2\pi)^{-1/2} e^N N^{-5/2} 
\sim  e^{\mu_0 N} N^{\gamma-3} \: ,
\label{canapprox}
\end{equation}
where the last step is due to the comparison with 
(\ref{susc}). The critical value of the chemical 
potential and the susceptibility exponent take 
the values $\mu_{0}=1$ and $\gamma=1/2$, respectively.
It turns out that the value $\gamma=1/2$ is a generic 
one. It does not change for a wide class of 
the weights. In the section about the universality classes
and singularity types two other universality classes of
branched polymer models with $\gamma \neq 1/2$ 
will be discussed.  

\begin{figure}[htbp]
\centerline{\epsfig{file=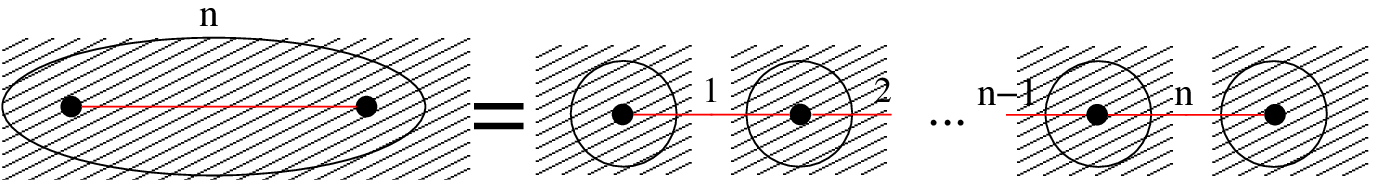, 
bburx=384, bbury=51, bbllx=0, bblly=0, width=12.0cm, angle=0}}
\caption[Decomposition of the internal two-point function.]{Decomposition
of the internal two-point function ${\cal G}^{(2)} \left( \mu , n \right)$.}
\label{twopf}
\end{figure}

We will close this appendix with the 
calculation of the internal two-point function.
Similarly as for the partition function, it is 
easier to work with the generating function.
Consider tree graphs which are weighted with the 
fugacity $g=e^{-\mu}$, and which
have two marked vertices separated by $n\geq 1$ links.
The generating function ${\cal G}^{(2)}\left(\mu, n \right)$ 
defined as a sum over all such trees corresponds to
the two-point function for the grand-canonical ensemble.
Figure (\ref{twopf}) illustrates the decomposition
of ${\cal G}^{(2)} \left( \mu ,n \right)$ into the 
generating functions $gV'(\Phi)$ and $gV''(\Phi)$ 
depicted in figure (\ref{phifig}). The decomposition
is unique, since the path connecting the 
marked vertices is unique. 
The two bubbles at the ends of the chain 
correspond to diagrams
of the generating function $gV'(\Phi)$, while
the $n-1$ ones in between to $gV''(\Phi)$. 
The decomposition leads to the following relation~:
\begin{equation}
{\cal G}^{(2)} \left( \mu , n \right) = e^{-\left(n+1\right)\mu} 
\left[ V' \left( \Phi \right) \right]^{2} \cdot 
\left[ V''\left( \Phi \right) \right]^{\left( n-1 \right)} \: .
\label{tpdef}
\end{equation}

We can also define the internal grand-canonical
two-point function for $n=0$. In this 
case the two marked vertices lie on top of each other. Thus
the two-point function reduces to the one-point function~:
${\cal G}^{(2)} \left( \mu , n=0 \right) = 
{\cal Z}^{(1)} = e^{-\mu} V(\Phi)$.

Relation (\ref{tpdef}) allows us to find an explicit dependence 
of the grand-canonical two-point function on $n$ and $\mu$,
if we first solve the self-consistency equation 
(\ref{gofphi}) for $\Phi(\mu)$.
We will be rather interested in the scaling behaviour
of the two-point function near the critical point.

Let us first illustrate the calculation of the two-point
function for the ensemble of trees having the natural
weight $W_T=1$. 
In this case $V'(\Phi) = V''(\Phi) = e^\Phi$,
where $\Phi(\mu)$ is a solution 
of equation (\ref{gofphi})~:
\begin{equation}
\mu = \Phi - \log (\Phi) \, .
\label{muphi}
\end{equation}
In this case the two-point function simplifies to~:
\begin{equation}
\label{trivtreegctp}
{\cal G}^{(2)} \left( \mu , n \right) = 
\exp \left( -(n+1)\left(\mu-\Phi(\mu)\right) \right) 
= \Phi^{n+1} \: .
\end{equation}
The inversion of equation (\ref{muphi}) for $\Phi$ gives
$\Phi = 1 - \sqrt{2}\sqrt{\Delta \mu}$, 
when $\Delta \mu = \mu \! - \! \mu_{0} 
= \mu\! - \! 1 \rightarrow 0^{+}$.
Thus the two-point function
can be approximated in this limit by~:
\begin{equation}
{\cal G}^{(2)} \left( \mu , n \right) \sim 
\exp \left( - \sqrt{2} (n+1) \sqrt{\Delta \mu} \right) \: .
\label{g22}
\end{equation}
We have neglected a term linear in $\Delta \mu$ in the exponent, 
because we are interested in the limit 
$\Delta \mu \rightarrow 0^{+}$ 
for which $\Delta \mu \ll \sqrt{\Delta \mu}$.
What matters in this limit is the leading term in $\Delta \mu$
which is related to the large $N$-behaviour of the underlying
canonical-ensemble~:
\begin{equation}
{\cal G}^{(2)} \left( \mu , n \right) = 
\sum_N {\cal G}^{(2)}_N\left(n\right) e^{-\mu N} =
\sum_N e^{-\mu_{0} N} {\cal G}^{(2)}_N\left(n\right) 
e^{-\Delta \mu N} \:,
\label{lt}
\end{equation}
where ${\cal G}^{(2)}_N(n)$ is the two-point function for the
canonical ensemble for trees of size $N$. In the last formula
we split into $\mu = \mu_{0} + \Delta \mu$, where
$\mu_{0}$ is the critical value of $\mu$ at which the
partition function is singular.
The leading terms in $\Delta \mu$ 
of ${\cal G}^{(2)} \left( \mu , n \right)$ are responsible
for the scaling behaviour 
while the next-to-leading ones for finite-size corrections. 

Formula (\ref{lt}) is a discrete Laplace transform.
Since we are interested in the large $N$ behaviour 
of ${\cal G}^{(2)}_N(n)$, we can substitute the discrete 
by a continuous Laplace transform. 
The inverse transform then yields~:
\begin{equation}
{\cal G}^{(2)}_N\left(n\right) = e^{+\mu_{0} N} 
\frac{1}{2\pi i} \int^{\Delta \mu_r + i\infty}_{\Delta \mu_r - i\infty}
{\rm d} \Delta \mu \ {\cal G}^{(2)} \left(\mu , n \right) 
e^{\Delta \mu N} \: .
\label{GIT}
\end{equation}

In particular, for the case discussed here 
(see(\ref{g22})) the exact result reads~:
\begin{equation}
\label{trivtreectp}
{\cal G}^{(2)}_N\left(n\right) = (2\pi)^{-1/2} e^{N} N^{-3/2} 
(n\!+\! 1) \exp \left( - \frac{(n\!+\! 1)^2}{2N} \right) \: .  
\end{equation}
We inserted $\mu_{0}=1$ in the last formula. 
The normalized two-point function can be approximated by~:
\begin{equation}
g^{(2)}_N(n) = 
\frac{{\cal G}^{(2)}_N\left(n\right)}{\sum_{n'} 
{\cal G}^{(2)}_N\left(n'\right)}
\simeq \frac{c n}{N} e^{-c n^2/2N} \: .
\label{ggn2}
\end{equation}
The constant $c$ in the last formula is equal $1$. We
displayed this constant, because the same formula holds
for generic trees in general, but with a constant
which depends on the choice of the weights.

We used two approximations in the last formula.
We substituted $n+1$ by $n$. The difference
between the function with $n$ and $n+1$
disappears in the large $N$-limit. Secondly,
the numerator and the denominator in the normalized
two-point function (\ref{ggn2}) have a common part
$(2\pi)^{-1/2} e^{N} N^{-3/2}$ which does not depend on $n$.
It cancels out. The remaining part~: $c/N$ is a normalization
constant ensuring $\sum_ng^{(2)}_N(n)=1$. 
For finite $N$ there will be some corrections
to the normalization constant $c/N$, 
but these disappear exponentially
in the large $N$ limit.


\begin{thebibliography}{99}
\bibitem{foot2} In the same way one can
show that all moments of the two-point function
scale uniformly for large $N$~:
\[ \int X^{K} G_N^{(2)}(X) \sim N^{K/4} \: , \]
which means that the two-point function is 
indeed a function
of the scaling argument $X/N^{1/4}$ for large $N$.
\bibitem{foot3} For the simulation
we used the exact form of $G^{(2)}_{N}(X)$ with the 
constants $c$ and $\sigma$ set equal to one~: \\
\[
G^{(2)}_{N}(X)=\sum_{n=1}^{N-1} 
\frac{e^{-\frac{X^2}{2n}}}{\left( 2 \pi n \right)^{D/2}} 
( n+1 ) e^{- \frac{(n+1)^2}{2 N}} \; . \]
\bibitem{foot4} If all moments of $f(x)$
exist, the size of the central region grows faster with $n$.
For example as $X_* \sim n^{3/4}$ for an even distribution. 
Therefore the gaussian regime sets in faster.
\end{thebibliography}
\end{document}